\documentclass[11pt]{article}
\pdfoutput=1

\usepackage[numbers,sort&compress]{natbib}
\usepackage{euscript}
\usepackage{amssymb}
\usepackage{amsfonts}
\usepackage{amsbsy}
\usepackage{epsfig}
\usepackage{amsthm}
\usepackage{amscd}
\usepackage{amstext}
\usepackage{verbatim}
\usepackage{amsmath}
\usepackage{cancel}
\usepackage{capt-of}

\usepackage[pdftex]{hyperref}

\def\ben{\begin{equation}}
\def\een{\end{equation}}

\def\be{\begin{equation}}
\def\ee{\end{equation}}
\def\beq{\begin{equation}}
\def\eeq{\end{equation}}
\def\ba{\begin{array}}
\def\ea{\end{array}}

\def\dalemb#1#2{{\vbox{\hrule height .#2pt
       \hbox{\vrule width.#2pt height#1pt \kern#1pt
               \vrule width.#2pt}
       \hrule height.#2pt}}}
\def\square{\mathord{\dalemb{6.8}{7}\hbox{\hskip1pt}}}
\newcommand{\bea}{\begin{eqnarray}}
\newcommand{\eea}{\end{eqnarray}}

\begin{document}

\begin{center}

{ \Large {\bf
Shear Viscosity to Entropy Density Ratio in Higher Derivative Gravity with Momentum Dissipation }}

\vspace{1cm}

{\bf Yi-Li Wang}, {~~~\bf Xian-Hui Ge}

\vspace{1cm}

{\small
{\it Department of Physics, Shanghai University,
Shanghai, 200444, China\\
Shanghai Key Laboratory of High Temperature Superconductors, Shanghai University, Shanghai 200444, P.R. China\\ }}

\vspace{1.6cm}

\end{center}

\begin{abstract}
    We investigate $\eta/s$ in linear scalar fields modified Gauss-Bonnet theory that breaks translation invariance.  We first calculate $\eta/s$ both analytically and numerically and show its relationship with temperature in log-log plot. Our results show that $\eta/s\sim T^2$ at low temperatures. The causality is also considered in this work. We then find that causality violation still happens in the presence of the linear scalar field and we suggest there is a Gauss-Bonnet coupling  dependent lower limit for the effective mass of the graviton. If the effective mass of the graviton is big enough, then there will be no causality violation and hence no constraints for the Gauss-Bonnet coupling.
\end{abstract}
\textbf{ } \\
\textbf{ }
\pagebreak
\setcounter{page}{1}

\section{Introduction}

Over the past decade, the anti-de Sitter (AdS)/ conformal field theory (CFT) correspondence has allowed us to stimulate much work on the dynamics of strongly coupled gauge theories. One of the most attractive among them is the ratio of the shear viscosity to the entropy density, as the black holes naturally have a good feature of thermodynamic properties \cite{hawking}, for example, entropy and temperature, which was analyzed in \cite{jacob}. For many strongly interacting quantum field theories, the dual holographic description of whom involves black holes in AdS space, we have a universal bound reading\cite{dtson, ao, pk, ab, son}
\begin{equation}\label{bound1}
\frac{\eta}{s}\ge \frac{1}{4\pi}\frac{\hbar}{k_B}.
\end{equation}

This is the well-known Kovtun-Starinets-Son (KSS) bound. It is well supported by the fact that all known fluids whose $\eta/s$ has been measured satisfy the bound \cite{oriol} including the quark-gluon plasma created at the Relativistic Heavy-Ion Collider (RHIC) \cite{derek, paul, song, rom, dusling}. In\cite{saso} and \cite{felix}, it has been further conjectured that in the dual theory the bound (\ref{bound1}) is related to the minimum entropy production of the black hole. In most cases, one can obtain the shear viscosity with a retarded Green function through the Kubo formula which takes the form
\begin{equation}\label{kubo}
\eta=\lim_{\omega\to0}\frac{1}{2\omega}\int dt d\textbf{x}e^{i\omega t}<[T_{xy}(t,\textbf{x}),T_{xy}(0,\textbf{0})]>,
\end{equation}
where $T_{xy}$ is the $xy$ component of the stress-tensor. There provides a prescription in \cite{dam} that enables one to calculate the Green function from gravity in Minkowski spacetime.

Interestingly, from the point of view of AdS/CFT, generic small corrections to Einstein gravity could obviously violate the bound (\ref{bound1}) because of the fact that the KSS bound is saturated by Einstein gravity \cite{liu}. For type-$\uppercase\expandafter{\romannumeral2}$B supergravity, \cite{mb} and \cite{kats} demonstrate that the viscosity bound is challenged when considering the stringy correction to the low-energy effective action. Furthermore, the causality would be violated as well in \cite{liu}. In \cite{ge1}, the charge dependence of $\eta/s$ has been computed for Gauss-Bonnet theory, after which the analysis of causality violation and instability of charged black brane has also been done. Ref. \cite{musso} makes a careful analysis of the heat and thermal diffusion constants in models with momentum dissipation. Different from the simplest holographic model with a bulk massive graviton, the model added a dilaton field shows a linear-in-$T$ resistivity and a constant electric susceptibility \cite{musso}. The work of \cite{oriol} also adds the graviton mass term and illustrates the impact of it on the viscosity. Ref. \cite{oriol} additionally hold the point of view that the viscoelastic nature of the mechanical response in materials as the physical reason is responsible for the bound violation. Different from the slight violation caused by higher derivative gravity, great impact can result from anisotropic models for strongly coupled $N=4$ Super-Yang-Mills plasma derived from type-$\uppercase\expandafter{\romannumeral2}$B supergravity as another example in Einstein gravity \cite{rebhan,Roy,mamo,glns,jain,Sadeghi:2015vaa,lu}, whose anisotropic background is based on the solution given and analyzed in \cite{dm} and \cite{diego}. Ref. \cite{ge3} gives a precise explanation of the causes of shear viscosity violation in anisotropic black branes. It is claimed that the equation of motion for the metric fluctuations failed to be written in the same form as the massless Klein-Gordon equation $\square h^{y}_{x}=0$. Since in higher derivative gravity, the Gauss-Bonnet coupling constant is constrained due to the causality \cite{cai,gltw}, the bound violation will not be so great as that in anisotropic black branes.

More recently, much progress has been made in the research on $\eta/s$. Further research has been done in \cite{yt} on gravity/fluid duality in an anisotropic gravitational system with Petrov-like boundary condition. Refs. \cite{ge1}, \cite{ge2} and \cite{ge4} have showed the upper bound of the Gauss-Bonnet coupling constant. In \cite{napat}, however, the standard hydrodynamic formula above is no longer valid if the scalar fields are included in the constitutive relation. Ref. \cite{napat} argues that the shear viscosity $\eta$ may not be the only cause of entropy production according to the modified constitutive relation. Considering more generic cases, the shear viscosity is evaluated with Gauss-Bonnet corrections in an anisotropic system, which comes after the calculation for the black brane solution in such a background \cite{viktor}.

Focusing on four-dimensional bulk spacetimes where the mass of the fluctuations of the metric components $\delta g_{xy}$ does not vanish, the authors of \cite{santos} have found that the viscosity to entropy density ratio $\eta/s$ would tend to a constant at low temperatures. It was suggested that the ratio of shear viscosity to entropy density is equal to the logarithmic increase of the entropy production per `Planckian time' \cite{santos, nature}. Suppose there exists a scale $\Delta$ in the zero temperature IR theory, it is reasonable to consider the presence of a different temperature-independent source $\delta g_{xy}^{0}=t\Delta$ that bounds the entropy production, in \cite{santos}, it is given that
\begin{equation}\label{santos}
\frac{\eta}{s}\gtrsim \left(\frac{T}{\Delta}\right)^2,
\,\frac{T}{\Delta}\to 0.
\end{equation}

This new bound, however, is violated in \cite{ly}. Given the bound for viscosity that reads $\eta/s\sim T^{\kappa}$, it has been found that $\kappa$ could be larger than 2 in hyperscaling violating geometry with lattice structure and a conjecture is made that the bound violation exists because of the  behaviours of entanglement entropy in  hyperscaling violating theories \cite{ly}. Different natures of ground states lead to different bounds.

When there exists momentum dissipation, $\eta/s$ will violate the KSS bound and Gauss-Bonnet gravity has its own special bound for $\eta/s$ \cite{mb}, which reads $4\pi\eta/s=(1-4\lambda)$, where $\lambda$ is the Gauss-Bonnet coupling constant. There exists a constraint that $\lambda<0.09$ because of the causality \cite{mb}. We would like to find out whether the viscosity bound violation and the constraint for $\lambda$ will change if the effective mass of perturbations of the metric components is not vanishing. It is also our interest to find the answer to such a question: Is the new viscosity bound (\ref{santos}) also true for Gauss-Bonnet gravity?

In this paper, we add both effective mass and Gauss-Bonnet term, computing viscosity to entropy density $\eta/s$ in higher derivative gravity with linear scalar fields adding to Gauss-Bonnet term. Causality structure analysis is also included. The body of this work involves two main parts. In section 2, we are going to evaluate the viscosity to entropy density ratio with the weaker horizon formula given in \cite{santos}.
The calculation is consist of both analytic and numerical solutions. In particular, we obtain the analytic results with expansions for small and large $\alpha/T$ separately. It turns out that our results are in good agreement with the bound (\ref{santos}). In section 3, we would like to find whether there exists causality violation in our cases. We find that there exists a constraint for the effective mass, and the constraint is a function of $\lambda$. The conclusions and discussions are provided in section 4.

\section{Viscosity to entropy density ratio}
\subsection{Background}

We start with the black brane solutions in Einstein-Maxwell-Gauss-Bonnet gravity with linear scalar fields. The five-dimensional action of gravity with three scalar fields is
\begin{equation}
S=\frac{1}{2\kappa^2}\int_{M}d^5x\sqrt{-g}\left(R-2\Lambda+\frac{\lambda}{2}\mathcal{L}_{GB}-\frac{1}{2}\sum_{i=1}^{3}(\partial\phi_i)^2-\frac{1}{4}F_{\mu\nu}F^{\mu\nu}\right),
\end{equation}
where the cosmological constant $\Lambda=-6$, the five-dimensional coupling $2\kappa^2=16\pi G_5$, and $\lambda$ represents Gauss-Bonnet coupling constant. In addition, following the conventions of curvatures as in \cite{sc}, the Gauss-Bonnet term reads
\begin{equation}
\mathcal{L}_{GB}=\left(R_{\mu\nu\rho\sigma}R^{\mu\nu\rho\sigma}-4R_{\mu\nu}R^{\mu\nu}+R^2\right).
\end{equation}

The equations of motion take the form
\begin{equation*}
\nabla_{\mu}F^{\mu\nu}=0,
\end{equation*}
\begin{equation*}
\nabla_{\mu}\nabla^{\mu}\phi_{i}=0,
\end{equation*}
\begin{eqnarray}\label{moeqn}
R_{\mu\nu}-\frac{1}{2}g_{\mu\nu}\left(R+12+\frac{\lambda}{2}(R^2-4R_{\rho\sigma}R^{\rho\sigma}+R_{\lambda\rho\sigma\tau}R^{\lambda\rho\sigma\tau})\right)\nonumber\\
+\frac{\lambda}{2}\left(2RR_{\mu\nu}-4R_{\mu\rho}R_{\nu}^{\ \rho}-4R_{\mu\rho\nu\sigma}R^{\rho\sigma}+2R_{\mu\rho\sigma\lambda}R_{\nu}^{\ \rho\sigma\lambda}\right)\nonumber\\
-\sum_{i=1}^{3}\left(\frac{1}{2}\partial_{\mu}\phi_{i}-\frac{g_{\mu\nu}}{4}\left(\partial\phi_i\right)^2\right)
-\frac{1}{2}\left(F_{\mu\lambda}F_{\nu}^{\ \lambda}-\frac{g_{\mu\nu}}{4}F_{\lambda\rho}F^{\lambda\rho}\right)=0.
\end{eqnarray}

Now we are going the find the viscosity to entropy density ratio with the following metric,
\begin{equation}
ds^2=\frac{r_{+}\left(-f(u)N^{2}dt^{2}+d\vec{x}^{2}+2h(t,z,u)dxdy\right)}{u}+\frac{du^2}{4u^2f(u)},
\end{equation}
 with $N$ found to be \cite{ge1}
\begin{equation}
N^2=\frac{1}{2}\left(1+\sqrt{1-4\lambda}\right),
\end{equation}
and
\begin{equation}
\phi_{i}=\alpha x_i, \ i=1,2,3.
\end{equation}
Simply quoting the results in \cite{cl,cai02} and taking the charge parameter $a$ to be zero, we have
\begin{equation}
f(u)=\frac{1}{2\lambda}\left(1-\sqrt{1-\lambda(1-u)(4+4u-\beta^2u)}\right),
\end{equation}
where we have performed a change of coordinates $u=\left(r_{+}/r\right)^2$ and set $\beta=\alpha/r_{+}$. In general, the metric fluctuation takes the form
\begin{equation}\label{per}
\left(\delta g\right)^{x}_{\ y}=h(u)e^{-i\omega t+ikz}.
\end{equation}

The Kubo formula (\ref{kubo}) provides us a clear definition for the shear viscosity in a relativistic quantum field theory. One can obtain $\eta/s$ from horizon data when in a translation invariant background \cite{nabil}. However, when the translation invariance is broken, the methods will not apply and a new formula is needed. We shall quote the 'weaker horizon formula' for $\eta/s$ in \cite{santos}, whose derivation has been appeared in \cite{lucas} and \cite{rad}. In cases where the perturbation is massive, we have
\begin{equation}\label{etas}
\frac{\eta}{s}=\frac{1}{4\pi}h_{0}(1)^2.
\end{equation}
According to the Kubo Formula (\ref{kubo}), $\omega$ in equation (\ref{per}) should be taken to be zero and thus we have the $h_0(u)$ in equation (\ref{etas}), which is the solution to the equation at zero frequency.

In our cases, the  equation(\ref{moeqn}) yields the equation for $h(u)$ after we change the coordinates
\begin{eqnarray}\label{eqnh}
0=h''(u)+\frac{g'(u)}{g(u)}h'(u)+\frac{\omega^2}{4uf^2(u)N^2r_{+}^{2}}h(u)\nonumber\\
-\frac{k^{2}\left(1-2\lambda u^2\left(2u(u^{-1}f(u))''+3(u^{-1}f(u))'\right)\right)}{4r_{+}^2uf(u)\left(1+2\lambda u^2(u^{-1}f(u))'\right)}
-\frac{\beta^2}{8u^2g(u)}h(u),
\end{eqnarray}
where $g(u)$ is given by
\begin{equation}
g(u)=u^{-1}f(u)\left[1+2\lambda u^2(u^{-1}f(u))'\right].
\end{equation}
and the related Hawking temperature is
\begin{equation}\label{tt}
T=\frac{\alpha}{\pi}\left(\frac{1}{\beta}-\frac{\beta}{8}\right).
\end{equation}
Since one can take the momentum to be zero($k=0$) and the solution for $h(r)$ that we need is the solution at zero frequency($\omega=0$), the equation (\ref{eqnh}) now takes the form
\begin{equation}\label{eqnh2}
h''(u)+\frac{g'(u)}{g(u)}h'(u)-\frac{\beta^2}{8u^2g(u)}h(u)=0,
\end{equation}
where we drop the subscript and call $h_0(u)$ as $h(u)$. Here $'$ denotes the differentiation with respect to $u$. It is obvious that the CFT is perturbed by the slope of the axion source $\alpha$ and the temperature $T$. When considering the higher derivative gravity, the viscosity bound is given \cite{mb}
\begin{equation}\label{bound2}
\frac{\eta}{s}\ge\frac{1}{4\pi}(1-4\lambda).
\end{equation}

As we have linear scalar fields, the bounds (\ref{bound2}) will probably be violated and there will be some changes. Now we are going to solve the equation analytically respectively in low and high temperature expansions. We will give the numerical results as well.

\subsection{High temperature expansion}

First we calculate the high temperature expansion, which means $\beta^{2}\sim0$. We expand the function $h(u)$ as

\begin{equation}\label{expansion}
h(u)=\sum_{i=0}^{+\infty}\beta^{2i}h_{2i}(u),
\end{equation}
and we expand the equation of motion for $h(u)$ in $\beta^2$. One could find that to first order, where $i$ is taken to be zero, the equation (\ref{eqnh2}) becomes

\begin{equation}
\left[g(u)h_{0}^{'}(u)\right]'=0.
\end{equation}
It vindicates that
\begin{equation}
g(u)h_{0}^{'}(u)=C_{1},
\end{equation}
where $C_{1}$ is an arbitrary constant.

Now one could claim that $C_1$ must be zero as $g(u)$ goes to zero at the horizon, which could be implied by the regularity of $h(u)$ at the horizon. Therefore, one finds the differentiation of $h_0(u)$ with respect to $u$ to be zero and thus
\begin{equation}\label{h0}
h_{0}(u)=C=\sqrt{1-4\lambda}.
\end{equation}
According to equation (\ref{bound2}), we choose $C$ to be $\sqrt{1-4\lambda}$ since it is the only solution satisfying the boundary conditions \cite{mb}.

At the second order, one obtains
\begin{eqnarray}\label{eqh2}
0=-\sqrt{1-4\lambda}-\frac{1}{\lambda(1-4\lambda+4u^{2}\lambda)^{\frac{3}{2}}}\left((4-4\sqrt{1-4\lambda+4u^{2}\lambda}\right.\nonumber\\
+\lambda(16u^{2}(-1+4\lambda)(-2+\sqrt{1-4\lambda+4u^{2}\lambda})\nonumber\\
-32(-1+2\lambda)(-1+\sqrt{1-4\lambda+4u^{2}\lambda})))h_{2}'(u)\nonumber\\
-\left.2u(2-8\lambda)(-1+4\lambda-4u^{2}\lambda)(-1+\sqrt{1-4\lambda+4u^{2}\lambda})h_{2}''(u)\right),
\end{eqnarray}
with the boundary conditions reading
\begin{equation*}
h_{2}(0)=0,
\end{equation*}
\begin{equation}
h_{2}'(1)=-\frac{1}{16\sqrt{1-4\lambda}}.
\end{equation}
To solve the equation, letting $U=\sqrt{1+4(-1+u^2)\lambda}$, one finds the equation (\ref{eqh2}) now becomes
\begin{eqnarray}
0=-8(U-2)(-1+4\lambda)(-1+U^2+4\lambda)^{\frac{3}{2}}h_{2}'(U)\nonumber\\
+U\left(-U^{3}\sqrt{\lambda(1-4\lambda)}+8(U-1)(-1+4\lambda)(-1+U^2+4\lambda)^{\frac{3}{2}}h_{2}''(U)\right).
\end{eqnarray}
Solve the equation and finally the solution for $h_2(u)$ is written as
\begin{eqnarray}
h_2(u)=\frac{1}{16\sqrt{1-4\lambda}}\left(\sqrt{1-4\lambda}+4u\lambda-2u^2\lambda-\sqrt{1+4(u^2-1)\lambda}-\sqrt{\lambda}\log(1-4\lambda)\right.\nonumber\\
+\log(1-\sqrt{1-4\lambda}-4\lambda)-\log(1-4\lambda-4u\lambda)-\sqrt{1+4(-1+u^2)\lambda}\nonumber\\
+\left.2\sqrt{\lambda}\log(2u\sqrt{\lambda}+\sqrt{1+4(-1+u^2)\lambda})\right).
\end{eqnarray}
We have now found the function for $h(u)$ and we now move to the calculation for shear viscosity to entropy density ratio. According to the definition of $T$ in equation(\ref{tt}), one finds in the high temperature limit
\begin{equation}
T=\frac{\alpha}{\pi}\left(\frac{1}{\beta}-\frac{\beta}{8}\right)\sim\frac{\alpha}{\pi\beta},
\end{equation}
and thus,
\begin{equation}\label{beta1}
\beta^2=\left(\frac{1}{\pi}\right)^2\left(\frac{\alpha}{T}\right)^2.
\end{equation}
Using equation (\ref{beta1}) and (\ref{etas}), we obtain the result for the shear viscosity to entropy density ratio at high temperature expansion as follows:
\begin{eqnarray}\label{hight}
4\pi\frac{\eta}{s}=1-4\lambda+\frac{1}{8\pi^2}\left(-1+\sqrt{1-4\lambda}+2\lambda+2\sqrt{\lambda}\log(1+2\sqrt{\lambda})\right.\nonumber\\
-\left.\sqrt{\lambda}\log(1-4\lambda)+\log(1-\sqrt{1-4\lambda}-4\lambda)-\log(-8\lambda)\right)\left(\frac{\alpha}{T}\right)^2.
\end{eqnarray}
One finds that the coefficient of $(\alpha/T)^2$ term is always negative, so it is obvious that $\eta/s$ would decrease as the temperature goes down. This means the KSS bound is violated.

\subsection{Low temperature expansion}
Let us now continue to solve the equation in the low temperature expansion. Letting $T=0$, one finds
\begin{equation}
T=\frac{\alpha}{\pi}\left(\frac{1}{\beta}-\frac{\beta}{8}\right)=0.
\end{equation}
Here one obtains $\beta^2=8$. Hence when dealing with the low temperature expansion, one could expand $h(u)$ around $\beta^2\sim8$ as
\begin{equation}
h(u)=\sum_{i=0}^{+\infty}\left(\beta^2-8\right)^{i}h_{i}(u).
\end{equation}
Expanding the equation of motion (\ref{eqnh2}) at $\beta^2\sim8$, one could get the equations for $h_0(u)$ and $h_1(u)$. As the concrete forms of the equations are too cumbersome, we would not press them here.
At zeroth order the boundary conditions are given by
\begin{equation*}
h_0(0)=\sqrt{1-4\lambda},
\end{equation*}
\begin{equation}\label{bd0}
h_0(1)=0.
\end{equation}
Then, at first order, one finds the boundary conditions read
\begin{equation*}
h_1(0)=0,
\end{equation*}
\begin{equation}\label{bd1}
h_1(1)=\frac{1}{4}h_0'(1).
\end{equation}
Similarly, with formula (\ref{tt}), one gets
\begin{equation}\label{beta2}
\beta^2-8=-8\sqrt{8}\pi\frac{T}{\alpha}.
\end{equation}
For $\eta/s$, we have
\begin{equation}
4\pi\frac{\eta}{s}=h^2(1)=\left(h_{0}(1)+(\beta^2-8)h_1(1)\right)^{2}=\frac{1}{16}\left(\beta^2-8\right)^{2}
\left(h_0'(1)\right)^2,
\end{equation}
where we use the boundary conditions (\ref{bd0}) and (\ref{bd1}), so all that we need is to find the value for $h_0'(1)$.
To this end, we solve the equation for $h_0(u)$ at the horizon ($u=1$) and in the region $u=0$, and then find the value for $h_0'(1)$ by matching these two solutions.

\subsubsection{Solution near $u=1$}
First, we expand $h(u)$ near the horizon in a Taylor series and call it $h_{01}(u)$:
\begin{equation}\label{taylor1}
h_{01}(u)=h_{01}'(u-1)+\frac{1}{2}h_{01}''(u)(u-1)^{2}.
\end{equation}
From the equation of motion for $h(u)$ (\ref{eqnh2}), $h_{01}¡¯¡¯(u)$ can be expressed as
\begin{eqnarray}\label{prime}
h_{01}''(u)=\frac{1}{u(1+4(u-1)\lambda)(-1+4(-1+u)^{2}\lambda)\left(-1+\sqrt{1-4(-1+u)^{2}\lambda}\right)}\nonumber\\
\cdot\left(2\lambda(1-4(-1+u)^{2}\lambda)^{\frac{3}{2}}h_{01}(u)\right.
-(-1+\sqrt{1-4(-1+u)^2\lambda}\nonumber\\
+16(-1+u)^2\lambda^2(-1+u+\sqrt{1-4(-1+u)^2\lambda})
-4\lambda(u(3-2\sqrt{1-4(-1+u)^2\lambda})\nonumber\\
+\left.u^2(-2+\sqrt{1-4(-1+u)^2\lambda})+2(-1+\sqrt{1-4(-1+u)^2\lambda})))h_{01}'(u)\right).
\end{eqnarray}
Taking the denominator of the expression (\ref{prime}) to be zero, one obtains
\begin{equation}\label{deno}
0=u(1+4(u-1)\lambda)(-1+4(-1+u)^{2}\lambda)\left(-1+\sqrt{1-4(-1+u)^{2}\lambda}\right).
\end{equation}
One would find that $u=1$ is a double root of this equation so it is necessary to use L'Hospital's rule twice if one wants to evaluate $h_{01}''(u)$ at $u=1$. Therefore, after the necessary procedure, we have
\begin{equation}\label{hpp}
h_{01}''(1)=-\frac{2}{5}\left(h_{01}'(1)+12\lambda h_{01}'(1)\right).
\end{equation}
Substituting equation (\ref{hpp}) into the expression (\ref{taylor1}), we obtain
\begin{equation}\label{taylor2}
h_{01}(u)=-\frac{1}{5}(u-1)(-6+u-12\lambda+12u\lambda)h_{01}'(1).
\end{equation}

\subsubsection{Solution near $u=0$}
Therefore, we have the expression for $h_0(u)$ at the horizon. We then continue to find the solution at $u=0$ called $h_{00}(u)$. Letting $u$ to be zero, one finds
\begin{equation}
f(u)=f(0)=\frac{1-\sqrt{1-4\lambda}}{2\lambda},
\end{equation}
and
\begin{equation}
g(u)=\frac{1}{2u\lambda}\left(1-\sqrt{1-4\lambda}\right)\sqrt{1-4\lambda}.
\end{equation}
Hence, according to the equation (\ref{eqnh2}) the equation of motion for $h_{00}(u)$  now reads,
\begin{equation}
0=8h_{00}(u)+\frac{4\left(1-\sqrt{1-4\lambda}\right)\sqrt{1-4\lambda}h_{00}'(u)}{\lambda}
-\frac{4u\left(1-\sqrt{1-4\lambda}\right)\sqrt{1-4\lambda}h_{00}''(u)}{\lambda},
\end{equation}
with the boundary condition given
\begin{equation}
h_{00}(0)=\sqrt{1-4\lambda}.
\end{equation}
One could obtain the solution for $h_{00}(u)$ written as
\begin{eqnarray}\label{h00}
h_{00}(u)=\frac{4u\lambda}{-1+\sqrt{1-4\lambda}+4\lambda}
\left(I_2(\frac{2\sqrt{2u\lambda}}{\sqrt{-1+\sqrt{1-4\lambda}+4\lambda}})\right.C\nonumber\\
+\left.\sqrt{1-4\lambda}K_2(\frac{2\sqrt{2u\lambda}}{\sqrt{-1+\sqrt{1-4\lambda}+4\lambda}})\right),
\end{eqnarray}
where $I_2$ is the modified Bessel function of the first kind, $K_2$ is the modified Bessel function of the second kind, and $C$ is a constant to be fixed.

\subsubsection{Matching}
We are now going to match the solutions (\ref{taylor2}) and (\ref{h00}) above at $u_m$, which is allowed to be an arbitrary number from 0 to 1. In order to connect them smoothly, it is required that
\begin{equation}
h_{00}\left(u_m\right)=h_{01}\left(u_m\right),
\end{equation}
\begin{equation}
h_{00}'\left(u_m\right)=h_{01}'\left(u_m\right).
\end{equation}
With these equations, one can easily find the exact expression of $h_{0}'(1)$ as a combination of hypergeometric functions of $\lambda$. We here simply give the direct result for $\eta/s$ at low temperature expansion:
\begin{equation}\label{lowt}
4\pi\frac{\eta}{s}=32\pi^2\left(h_0'(1)\right)^{2}\left(\frac{T}{\alpha}\right)^{2},
\end{equation}
where
\begin{eqnarray}
h_0'(1)=25(1-4\lambda)/\left(
16\left((u_m-1)(-6+u_m+12(u_m-1)\lambda)\frac{_0F_1(2;\frac{2\lambda u_m}{-1+4\lambda+\sqrt{1-4\lambda}})}{\Gamma(2)}\right.\right)\nonumber\\
\left.\left.+u_m(7-2u_m-24(u_m-1)\lambda)\frac{_0F_1(3;\frac{2\lambda u_m}{-1+4\lambda+\sqrt{1-4\lambda}})}{\Gamma(3)}\right)\right)^2,
\end{eqnarray}
and $_0F_1$ is the confluent hypergeometric limit function related to Bessel functions. From the equation above, one could find that the value for $h'_0(1)$ would go up with $u_m$ increasing. Indeed the $u_m$ that one chooses will not change the main features of the solution \cite{ruth}, and here we would like to take $u_m$ to be $1/10$. At $u_m=1/10$, we have
\begin{equation}
h_0'(1)=-\frac{500\sqrt{1-4\lambda}}
{9(59+108\lambda)_0F_1(;2 ;\frac{\lambda}{5(-1+4\lambda+\sqrt{1-4\lambda})})
+2(17+54\lambda)_0F_1(;3 ;\frac{\lambda}{5(-1+4\lambda+\sqrt{1-4\lambda})})}.
\end{equation}

We combine the results for high temperature expansion and low temperature expansion into Figure 1 in a log-log plot. We choose $4\pi\eta/(1-4\lambda)s$ as the vertical axis and $\alpha/T$ as the horizontal axis. The perturbative results for small $\alpha/T$ is shown as red dashed line, while the results close to extremality is shown as the green line.
\begin{figure}[h]
\centering
\includegraphics[scale=1.4]{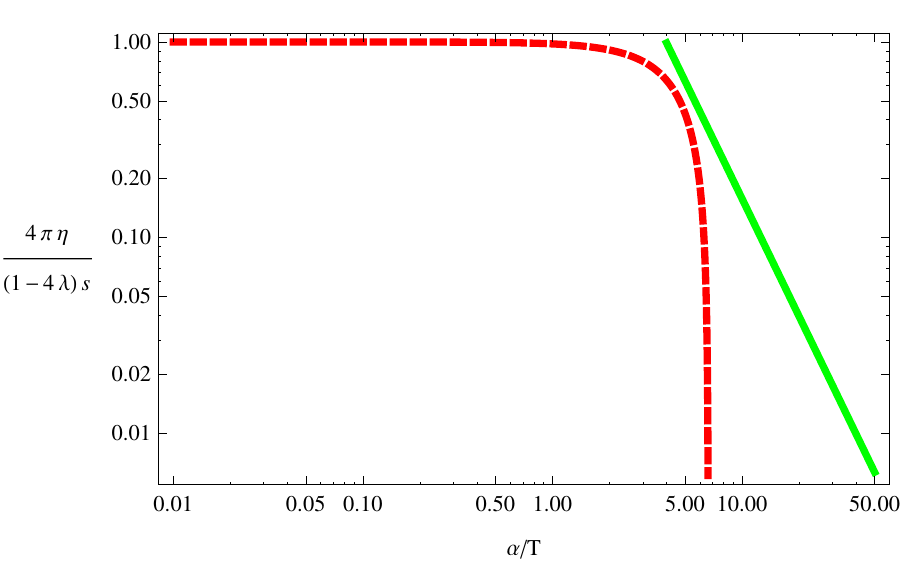}\hspace{0.7cm}
\caption{ Log-log plot of $4\pi\eta/s$ as a function of $\alpha/T$, in a background with higher derivative Gauss-Bonnet gravity. The dotted red line represents the analytical results at small $\alpha/T$, and the green one shows that for large $\alpha/T$. Here we set $\lambda=0.1$ and $u_m=1/10$.}
\end{figure}

\subsection{Numerical results}

With numerical methods, one can solve the equation (\ref{eqnh2}) for any value of $\beta$. At the boundary, we still require $h(0)=\sqrt{1-4\lambda}$ and at the horizon, the equation of motion yields
\begin{equation}
h'(1)=\frac{\beta^2h(1)}{\left(\beta^2-8\right)\left(2+(\beta^2-8)\lambda\right)}.
\end{equation}
As shown in Figure 2, we also give the chart in log-log plot for the numerical solutions of the equation in three dashed lines, with no less than 500 points on the each locus. The blue, red and green lines (from top to bottom) correspond to $\lambda=-0.1, 0.08, 0.15$ respectively.
\begin{figure}[h]
\center{
\includegraphics[scale=1.4]{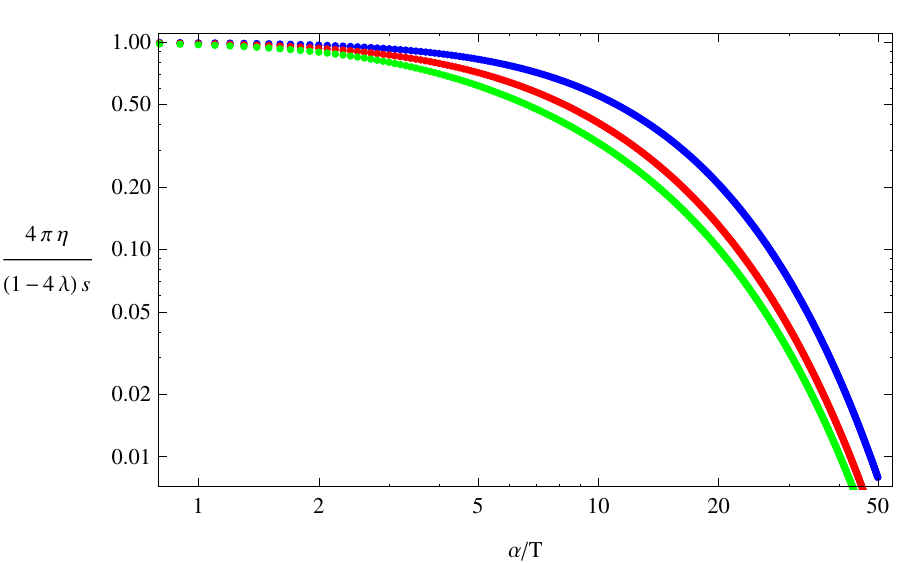}\hspace{0.7cm}
\caption{Log-log plot of the numerical results for the equation (\ref{eqnh2}). Here we set $\lambda=-0.1$, 0.08, and 0.15. The blue line represents the results for $\lambda=-0.1$, the red disks are data for $\lambda=0.08$, and the green ones are the numerical data obtained when $\lambda$ is set to be 0.15.} }
\end{figure}

\section{Causality}
It has been mentioned that one could find that the causality could be violated when introducing Gauss-Bonnet terms \cite{ge1, liu}. Now let us here  continue to analyze whether there is causality violation if we add linear scalar fields together with Gauss-Bonnet term.

We rewrite the wave function as
\begin{equation}
h(x,u)=e^{-i\omega t+ikz+ik_{u}u}.
\end{equation}
Then we take large momenta limit, which means $k^{\mu}\to\infty$, and the equation (\ref{eqnh}) can be rewritten into
\begin{eqnarray}\label{qnm}
0=4u^{2}f(u)k_{u}^2+\frac{c_{g}^{2}u}{r_{+}^{2}N^{2}f(u)}k^{2}-\frac{u}{r_{+}^{2}N^{2}f(u)}\omega^{2}\\
+\frac{f(u)}{2g(u)}\beta^2-\frac{4u^{2}f(u)g'(u)}{g(u)}(ik_{u}).\nonumber
\end{eqnarray}
One could simplify the equation into
\begin{equation}\label{cau}
k_{\mu}k_{\nu}g^{\mu\nu}_{eff}\sim0,
\end{equation}
where the effective metric reads
\begin{equation}
ds^{2}_{eff}=g^{eff}_{\mu\nu}dx^{\mu}dx^{\nu}=\frac{N^{2}f(u)r_{+}^{2}}{u}\left(-dt^{2}+\frac{1}{c_{g}^{2}}dz^{2}\right)+\frac{1}{4u^{2}f(u)}du^2.
\end{equation}
Note that $N^2=\left(1+\sqrt{1-4\lambda}\right)/2$, and the terms of $ik_u$ and $\beta^2$ vanish as $k^{\mu}$ is taken to be infinity.
Here $c_g^2$ could indicate the local speed of graviton and it has the expression as follows:
\begin{equation}\label{cg}
c_{g}^{2}(u)=\frac{N^{2}f(u)\left(1-2\lambda u^{2}(2u(u^{-1}f(u))''+3(u^{-1}f(u))')\right)}
{1+2\lambda u^2(u^{-1}f(u))'}.
\end{equation}
Expanding $c_g^2$ at the boundary ($u=0$), one finds
\begin{equation}\label{first}
c_g^2-1=\frac{2\lambda}{1-\sqrt{1-4\lambda}}
\left(-\frac{\beta^2 u}{4\sqrt{1-4\lambda}}\right)+\mathcal{O}(u^2).
\end{equation}
It is reasonable to expand $c_g^2$ at the boundary only to the first order of $u$, since $u$ is taken to be a very small number. As the local speed of light of boundary CFT is specified to be unity ($c=1$), the local speed of graviton $c_g^2$ should be no larger than $1$. One could find that the term on the right-hand side would always be negative as long as $\lambda<1/4$ for any real number valued of $\beta$.

However, when the effective mass vanishes ($\beta^2=0$), the result will be different. The term to the first order of $u$ in equation (\ref{first}) will be vanishing as $\beta^2$ is set to be zero, so one has to expand $c_g^2$ to the second order which reads
\begin{equation}\label{second}
c_g^2-1=\frac{2\lambda}{1-\sqrt{1-4\lambda}}
\left(-\frac{((-5+4\sqrt{1-4\lambda}+20\lambda)(-16+64\lambda))u^2}
{16(1-4\lambda)^{\frac{5}{2}}}\right).
\end{equation}
As we have mentioned above, the local speed of graviton should be smaller than $1$. This yields the constraint for $\lambda$ from the equation (\ref{second}), say $\lambda<0.09$. Therefore, one recovers the results in \cite{ge1, liu}, claiming that $c_g^2$ is larger than $1$ in the regime $\lambda>0.09$ for either neutral or charged black holes.

\begin{figure}[h]
\center{
\includegraphics[scale=1.2]{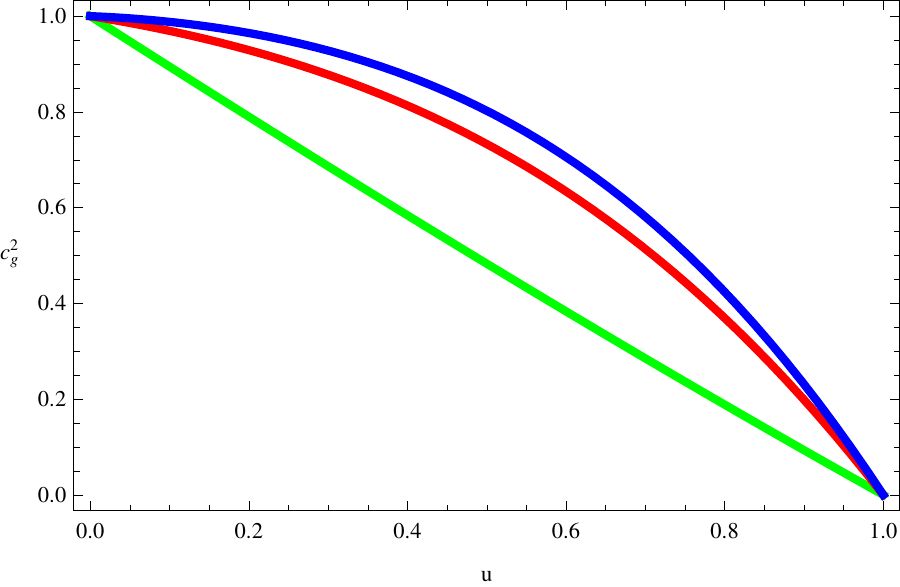}\hspace{0.7cm}
\caption{Diagram for the local speed of graviton as a function of $u$. Here $\lambda=0.05$. The bottom blue line corresponds to $\beta^2=1/4$, the middle red one represents $c_g^2$ for $\beta^2=1$, and the top green line illustrates the case where $\beta^2=4$.} }
\end{figure}
\begin{figure}[h]
\center{
\includegraphics[scale=1.2]{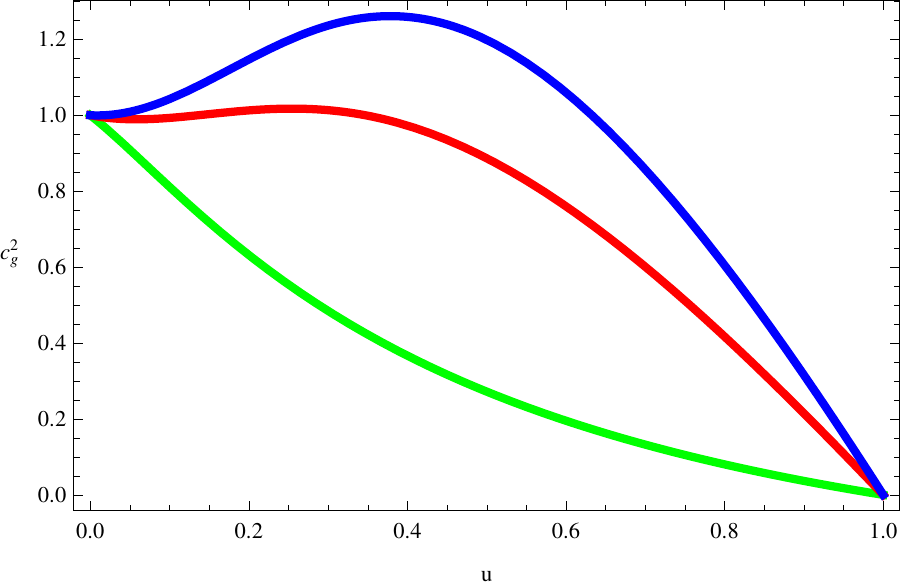}\hspace{0.7cm}
\caption{The local speed of graviton for $\lambda=0.2$. The blue, red and green lines (from bottom to top) are corresponding to $\beta^2=1/4, 1, 4$ respectively.} }
\end{figure}

A through study of the causality structure calls for the analysis of the bulk speed of the gravitons. Interestingly, it seems that $c_g^2$ in equation (\ref{cg}) could be greater than $1$ in the bulk when $\lambda>0.09$, which is illustrated in Figure 3 and Figure 4.
According to the analysis above, one confirms that when $\lambda<0.09$, there is no causality violation, no matter how small $\beta^2$ is. When one deals with the case in which $\lambda>0.09$, however, $c_g^2$ might be larger than $1$ because of the value chosen for $\beta^2$. To analyze the behavior of $c_g$, we would now study in the regime $\lambda>0.09$ and see how the causality is violated. Following the procedure in \cite{liu,ge1}, one could change the equation into equation (\ref{qnm}) for geodesic motion of gravitons,
\begin{equation}
\frac{dx^{\mu}}{ds}\frac{dx^{\nu}}{ds}g^{eff}_{\mu\nu}=0,
\end{equation}
after using the relation $\frac{dx^{\mu}}{ds}=g^{\mu\nu}_{eff}$. Furthermore, $\omega$ and $k$ are defined by
\begin{equation}
\omega=\left(\frac{dt}{ds}\right)\frac{f N^2r_+^2}{u}, \qquad k=\left(\frac{dz}{ds}\right)\frac{fN^2r_+^2}{u c_g^2}.
\end{equation}
As shown in Figure 4, starting at the boundary, the geodesic line of the gravitons would bounce back to the boundary. There is a turning point, say $u_0$, where $\omega^2/k^2=c_g^2(u_0)$.
Here $r_0$ corresponds to $u_0$ as the turning point. For a light-like geodesic line, one finds
\begin{equation}
\Delta t=2\int_{r_0}^{\infty}\frac{dt}{d\tilde{s}}\frac{\tilde{s}}{dr}dr=\frac{2}{N}\int_{r_0}^{\infty}dr\frac{\zeta}{f(r)\sqrt{\zeta^2-c_g^2}},
\end{equation}
\begin{equation}
\Delta z=2\int_{r_0}^{\infty}\frac{dz}{d\tilde{s}}\frac{\tilde{s}}{dr}dr=\frac{2}{N}\int_{r_0}^{\infty}dr\frac{c_g^2}{f(r)\sqrt{\zeta^2-c_g^2}},
\end{equation}
where $\zeta^2=\omega^2/k^2$ and $\tilde{s}=ks/N$.
Call the maximum value that $c_g^2$ could reach as $c_{g,max}$. When $\frac{\Delta z}{\Delta t}>1$, there happens microcausality violation according to \cite{liu,ge1}. As $c_g(u_0)\to c_{g,max}$, one thus obtains
\begin{equation}
\frac{\Delta z}{\Delta t} \to c_{g,max}>1.
\end{equation}
Then we change the wave function into a Schr$\ddot{o}$dinger form,
\begin{equation}
-\frac{d^2\psi}{dr_*^2}+V(r(r_*))\psi=\left(\frac{\omega}{2r_+}\right)^2\psi,\qquad \frac{r_*}{r}=\frac{r_+^2}{N f(r)},
\end{equation}
where
\begin{equation*}
\psi=K(r)\phi, \qquad K(r)\equiv \sqrt{\frac{g(u)}{u^{-1}f(u)}}=1-\lambda\frac{r}{r_+}\frac{\partial(f(r)/r_+^2)}{\partial r}, \\
\end{equation*}
\begin{equation*}
V=k^2c_g^2+V_1(r),\qquad V_1(r)\equiv \frac{N^2}{r_+^2}\left[\left(f(r)\frac{\partial lnK(r)}{\partial r}\right)^2
+f(r)\frac{\partial}{\partial r}\left(f(r)\frac{\partial lnK(r)}{\partial r}\right)\right].
\end{equation*}
Here the Bohr-Sommerfeld quantization condition is applied and one has
\begin{equation}
\frac{k}{2r_+}\int^{0}_{r_{*0}}dr_*\sqrt{\zeta^2-c_g^2}=(n-\frac{1}{4})\pi,
\end{equation}
where $n$ is integer.
Following \cite{liu,ge1}, we finally get the group velocity of the graviton which takes the form
\begin{equation}
v_g=\frac{d\omega}{dk}=\frac{\Delta z}{\Delta t}.
\end{equation}
One would find that signals in the CFT may propagate outside of the light cone. Hence, we verify that in regime $\lambda>0.09$, the causality could still be violated. On the other hand, the theory with both Gauss-Bonnet term and effective mass is safe in regime $\lambda<0.09$. We also make a contour plot of $c_g^2$ labeled Figure $5$, and $u$ is set to be 0.5. From Figure 5, one finds there should also a new constraint for $\beta^2$ which reads $\beta^2\geq\beta_m^2$. One could see that $\beta^2_m$ is a function of $\lambda$, that is, the constraint for $\beta^2$ depends on the value of $\lambda$ that one chooses. As a result, $\beta_m^2$ should be written as $\beta_m^2(\lambda)$. Since the equation is rather complex and $c_g^2$ is a function of $u$, $\lambda$ and $\beta^2$, we are unable to give an explicit expression for the time being.
\begin{figure}[h]
\center{
\includegraphics[scale=0.8]{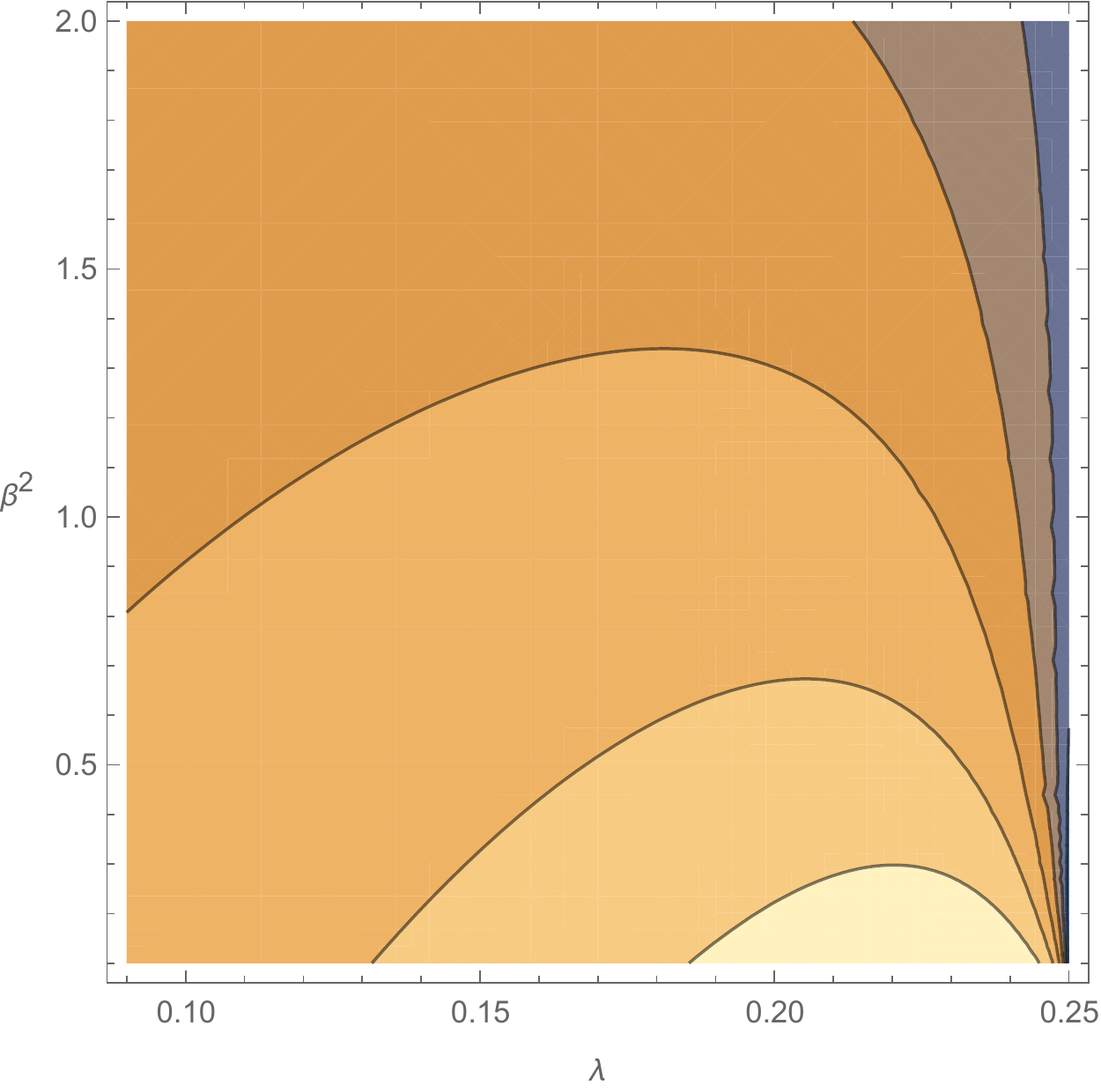}\hspace{0.7cm}
\caption{Contour plot of $c_g^2$ as a function of $\beta^2$ and $\lambda$. From bottom to top, the lines correspond to $c_g^2=1.2, 1.0, ..., 0.4$ respectively. Here $u$ is chosen to be $0.5$.} }
\end{figure}
\section{Conclusions and discussions}

In this work, we have investigated the shear viscosity to entropy density ratio with massive perturbation in the higher derivative gravity by adding the Gauss-Bonnet term. We quoted the formula in \cite{santos} which can be deduced from Kubo Formula to calculate $\eta/s$. The ratio is found to be less than $(1-4\lambda)/4\pi$, that is, it violates the bounds (\ref{bound2}). Our results illustrates that when the temperature is high, $\eta/s$ tends to be constant $(1-4\lambda)/4\pi$, while it decreases obviously when the temperature goes down. The results obtained (\ref{hight}) and (\ref{lowt}) indeed obey the formula (\ref{santos}). It shows that $\eta/s\sim T^2$ at the lowest temperature as $\Delta=\alpha$. Admittedly, the methods we use to calculate the low temperature expansion and the numerical solutions can be somehow improved to achieve more precise results, but the qualitative features will not change.

 When the effective mass of the graviton vanishes, we recover the results of \cite{liu,ge1}, which tells us that in the boundary CFT, causality violation will happen if $\lambda>0.09$. For nonvanishing effective mass of the graviton case, we then investigate further into the whole bulk on the geodesic motion of the graviton and find that there exists a lower limit $\beta_m^2(\lambda)$ for $\beta^2$ which depends on the Gauss-Bonnet constant $\lambda$. This means that causality imposes strong constraints on the value of the effective mass $\beta^2$ in the Gauss-Bonnet gravity theory.

Furthermore, as we have mentioned above, in this paper, we take the charge density $a$ to be zero. It is also our interests for future task to evaluate $\eta/s$ after adding charge together with linear scalar fields in Gauss-Bonnet gravity and to investigate whether there is causality violation. It is reasonable to expect that there exist more general bounds for transport coefficients and characterize the black hole.
\section*{Acknowledgements}

The authors were partly supported by NSFC, China (No.11375110)  and the Grant (No. 14DZ2260700) from  Shanghai Key Laboratory of High Temperature Superconductors..


\begin{thebibliography}{99}

 \bibitem{hawking}
S. W.Hawking, ``Particle creation by black holes,'' Commun. Math Phys. 43, 199 (1975).

\bibitem{jacob}
 Jacob D. Bekenstein, `` Black Holes and Entropy," Phys. Rev. D. 7, 2333 (1973).

  \bibitem{dtson}
P. K. Kovtun, D. T. Son and A. O. Starinets,
``Viscosity in strongly interacting quantum field theories from black hole physics,'' Phys. Rev. Lett. 94, 111601 (2005) [arXiv:hep-th/0405231].

\bibitem{ao}
G. Policastro, D.T. Son, A.O. Starinets, ``Shear viscosity of strongly coupled N=4 supersymmetric Yang-Mills plasma,'' Phys. Rev. Lett. 87, 081601 (2001) [arXiv:hep-th/0104066]


\bibitem{pk}
Pavel Kovtun, Dam T. Son and Andrei O. Starinets, ``Holography and hydrodynamics: diffusion on stretched horizons
,'' JHEP 0310. 064 (2003) [arXiv:hep-th/0309231].

\bibitem{ab}
 Alex Buchel and James T. Liu, ``Universality of the shear viscosity in supergravity,'' Phys. Rev. Lett. 93, 090602 (2004) [arXiv:hep-th/0311175]

\bibitem{son}
 D. T. Son and A. O. Starinets, ``Viscosity, Black Holes, and Quantum Field Theory,''  Ann. Rev. Nucl. Part. Sci. 57, 95 (2007) [arXiv:0704.0240[hep-th]].


\bibitem{oriol}
Lasma Alberte, Matteo Baggioli and Oriol Pujol$\grave{a}$s, ``Viscosity bound violation in holographic solids and the viscoelastic response,''
 [arXiv:1601.03384 [hep-th]].

\bibitem{derek}
 Derek Teaney, ``The Effect of Shear Viscosity on Spectra, Elliptic Flow, and HBT Radii,'' Phys. Rev. C 68, 134913 (2003) [arXiv:nucl-th/0301099].

\bibitem{paul}
 Paul Romatschke and Ulrike Romatschke, ``Viscosity Information from Relativistic Nuclear Collisions: How Perfect is the Fluid Observed at RHIC?,'' Phys. Rev. Lett. 99, 172301 (2007)[arXiv:0706.1522[nucl-th]].

\bibitem{song}
 Huichao Song and Ulrich Heinz, ``Suppression of elliptic flow in a minimally viscous quark-gluon plasma,'' Phys. Lett. B 658,279 (2008)[arXiv:0709.0742[nucl-th]].

\bibitem{rom}
 Paul Romatschke, ``Fluid turbulence and eddy viscosity in relativistic heavy-ion collisions,'' Prog. Theor. Phys. Suppl. 174, 137-144 (2008) [arXiv:0710.0016[nucl-th]].

\bibitem{dusling}
 K. Dusling and D. Teaney, ``Simulating elliptic flow with viscous hydrodynamics,'' Phys.Rev.C 77, 034905 (2008) [arXiv:0710.5932[nucl-th]].

\bibitem{saso}
 S. Grozdanov, A. O. Starinets, `` On the universal identity in second order hydrodynamics,'' JHEP 03, 007 (2015) [arXiv:1412.5685 [hep-th]].

\bibitem{felix}
 Felix M. Haehl, R. Loganayagam, Mukund Rangamani, `` Adiabatic hydrodynamics: The eightfold way to dissipation,'' JHEP 05, 060 (2015) [arXiv:1502.00636 [hep-th]].

\bibitem{dam}
Dam T. Son and Andrei O. Starinets, `` Minkowski-space correlators in Ads/CFT Correspondence: recipe and applications,'' JHEP 0209 (2002) 042 [arXiv:hep-th/0205051].

\bibitem{liu}
 Mauro Brigante, Hong Liu, Robert C. Myers, Stephen Shenker, and Sho Yaida, ``Viscosity Bound and Causality Violation,'' Phys. Rev. Lett. 100, 191601 (2008) [arXiv:0802.3318 [hep-th]].

\bibitem{mb}
 Mauro Brigante, Hong Liu, Robert C. Myers, Stephen Shenker, Sho Yaida, `` Viscosity Bound Violation in Higher Derivative Gravity,'' Phys. Rev. D. 77, 126006 (2008) [arXiv:0712.0805 [hep-th]].

\bibitem{kats}
 Yevgeny Kats and Pavel Petrov, `` Effect of curvature squared corrections in AdS on the viscosity of the dual gauge theory,'' JHEP 0901, 044 (2009) [arXiv:0712.0743 [hep-th]].

\bibitem{ge1}  Xian-Hui Ge, Yoshinori Matsuo, Fu-Wen Shu,Sang-Jin Sin and Takuya Tsukioka,`` Viscosity bound, causality violation and instability with stringy correction and charge,'' JHEP 10 (2008) 009.

\bibitem{musso}
Andrea Amoretti, Alessandro Braggio, Nicodemo Magnoli, Daniele Musso, ``Bounds on charge and heat diffusivities in momentum dissipating holography,''  JHEP 1507 (2015) 102 [arXiv:1411.6631[hep-th]].


\bibitem{rebhan}
Anton Rebhan and Dominik Steineder, ``Violation of the Holographic Viscosity Bound in a Strongly Coupled Anisotropic Plasma,'' Phys. Rev. Lett. 108 (2012)[arXiv:1110.6825].

\bibitem{mamo}
K.~A.~Mamo, ``Holographic RG flow of the shear viscosity to entropy density ratio in strongly coupled anisotropic plasma,''
 JHEP {\bf 1210}, 070 (2012). 

 \bibitem{glns} X. H. Ge, Y. Ling, C. Niu and S. J. Sin, ``Thermoelectric conductivities, shear viscosity, and stability in an anisotropic linear axion model
," Phys. Rev. D 92, 106005 (2015) [arXiv:1412.8346[hep-th]].

\bibitem{jain}S. Jain, R. Samanta and S. P. Trivedi, ``The shear viscosity in anisotropic phases,''
[arXiv:1506.01899[hep-th]].

\bibitem{Roy}A. Bhattacharyya, D. Roychowdhury, ``Viscosity bound for anisotropic superfluids in higher derivative gravity,''  JHEP 1503 (2015) 063 [arXiv:1410.3222 [hep-th]]

\bibitem{Sadeghi:2015vaa}
  M.~Sadeghi and S.~Parvizi,   ``Hydrodynamics of a Black Brane in Gauss-Bonnet Massive  Gravity,''   arXiv:1507.07183 [hep-th].

\bibitem{lu}
H. S. Liu, H. L¡§u, and C. N. Pope, ``Generalized Smarr formula and the viscosity bound for Einstein-Maxwell-dilaton black holes,''  Phys. Rev. D 92, 064014 (2015).

\bibitem{dm}
David Mateos, and Diego Trancanelli, ``The anisotropic N = 4 super Yang-Mills plasma and its instabilities,'' Phys. Rev. Lett. 107, 101601 (2011) [arXiv:1105.3472].

\bibitem{diego}
David Mateos and Diego Trancanelli, ``Thermodynamics and Instabilities of a Strongly Coupled Anisotropic Plasma,''JHEP 1107:054,2011 [arXiv:1106.1637].

\bibitem{ge3}
 Xian-Hui Ge,``Notes on shear viscosity bound violation in anisotropic models,'' Sci. China-Phys. Mech. Astron. 59, 630401 (2016) [arXiv:1510.06861 [hep-th]].

\bibitem{cai}
Rong-Gen Cai, Zhang-Yu Nie, Ya-Wen Sun, ``Shear Viscosity from Effective Couplings of Gravitons,'' Phys. Rev. D 78 (2008) [arXiv:0811.1665].

\bibitem{gltw}
X. H. Ge, Y. Ling, Y. Tian and X. N. Wu, ``Holographic RG Flows and Transport Coefficients in Einstein-Gauss-Bonnet-Maxwell Theory ,''JHEP 1201 117 (2012) 
 [ arXiv:1112.0627 [hep-th]].


\bibitem{yt}
Wen-Jian Pan, Yu Tian, and Xiao-Ning Wu, ``From Petrov-Einstein-Dilaton-Axion to Navier-Stokes equation in anisotropic model,'' Phys. Lett. B 752 (2016) [arXiv:1410.3222[hep-th]].

\bibitem{ge2}
X.-H. Ge and S.-J. Sin, ``Shear viscosity, instability and the upper bound of the Gauss-Bonnet coupling constant,'' JHEP 05 (2009) 051 [arXiv:0903.2527[hep-th]].

\bibitem{ge4}
X.H. Ge, S. Sin, S. Wu, and G. Yang  ``Shear viscosity and instability from third order Lovelock gravity,'' Phys. Rev. D, 80 (2009) 104019 [arXiv:0905.2675[hep-th]].

\bibitem{napat}
Piyabut Burikham and Napat Poovuttikul, ``Shear viscosity in holography and effective theory of transport without translational symmetry,'' [arXiv:1601.04624 [hep-th]].

\bibitem{viktor}
Viktor Jahnke, Anderson Seigo Misobuchi, and Diego Trancanelli, ``Holographic renormalization and anisotropic black branes in higher curvature gravity,''JHEP 01 (2015) 122 [arXiv:1411.5964 [hep-th]].

\bibitem{santos} Sean A. Hartnoll, David M. Ramirez and Jorge E. Santos, ``Entropy production, viscosity bounds and bumpy black holes,''
[arXiv:1601.02757 [hep-th]].

\bibitem{nature}
J. Zaanen, `` Superconductivity: Why the temperature is high,'' Nature, 430, 512 (2004).

\bibitem{ly}
Yi Ling, Zhuoyu Xian, and Zhenhua Zhou, ``Holographic Shear Viscosity in Hyperscaling Violating Theories without Translational Invariance,''
[arXiv:1605.03879 [hep-th]].

\bibitem{sc}
 S. Carroll, ``Spacetime and Geometry,''  Addision-Wesley Reading, MA, (2004).

\bibitem{cl}
  Long Cheng, Xian-Hui Ge and Zu-Yao Sun,  ``Thermoelectric DC conductivities with momentum dissipation from higher derivative gravity,'' JHEP 04 (2015) 135 [arXiv:1411.5452[hep-th]].
  \bibitem{cai02}  R.-G. Cai, ``Gauss-Bonnet black holes in AdS spaces," Phys. Rev. D 65 (2002) 084014
[hep-th/0109133]

\bibitem{nabil}
 Nabil Iqbal and Hong Liu, ``Universality of the hydrodynamic limit in AdS/CFT and the membrane paradigm,'' Phys. Rev. D 79, 025023 (2009) [arXiv:0809.3808[hep-th]].

 \bibitem{lucas}
 Andrew Lucas, ``Conductivity of a strange metal: from holography to memory functions,''  JHEP 1503, 071 (2015)  [arXiv:1501.05656[hep-th]].

\bibitem{rad}
 Richard A. Davison, Blaise Gout$\acute{e}$raux, and Sean A. Hartnoll, ``Incoherent transport in clean quantum critical metals,''  JHEP 1510, 112 (2015)[arXiv:1507.07137[hep-th]].


\bibitem{ruth}
Ruth Gregory, Sugumi Kanno, and Jiro Soda, ``Holographic Superconductors with Higher Curvature Corrections,'' JHEP 0910, 010 (2009) [arXiv:0907.3203[hep-th]].





\end{thebibliography}
\end{document}